\documentclass[aps,prb,twocolumn,showpacs]{revtex4}
\usepackage{graphicx}
\begin{document}

\title{Spin-Hall Conductivity and Pauli Susceptibility in the Presence of
Electron-Electron Interactions}

\author{Ol'ga V. Dimitrova}

\affiliation{L. D. Landau Institute for Theoretical Physics, Russian Academy 
of Sciences, Kosygina str. 2,  Moscow 119334, Russia}

\date{\today}

\begin{abstract}
We found the universal relationship between frequency-dependent spin-Hall 
conductivity 
and magnetic susceptibility in clean 2D electron systems with Rashba coupling:
$\sigma_{sH}(\Omega)= \frac{e}{(g\mu_B)^2m_b}\chi_\parallel(\Omega)$ 
in the presence of an
arbitrary two-particle spin-conserving interaction. 
We show that the Coulomb interaction renormalizes
the spin-Hall constant. The magnitude of the relative correction to 
$\sigma_{sH}$ is proportional to the Coulomb interaction parameter 
$e^2/\epsilon v_F\hbar$ and  does not depend on the strength of 
the Rashba coupling $\alpha$.
\end{abstract}

\pacs{72.25.-b}

\maketitle
Recently it has been proposed~\cite{Murakami} that a {\it dissipationless } 
spin current can be generated in response to an electric field in 
semiconductors with the spin-orbital interaction. For the case of an ideal 
two-dimensional 
(2D) electron gas with the Rashba coupling, Sinova et al.~\cite{Sinova} have 
found a spin-Hall current of the transverse ($z$) spin component as a response 
to an in-plane electric field $E_{\nu}$,
$j^z_\mu = \sigma_{sH} \epsilon_{\mu\nu} E_\nu$, 
with the ``universal'' spin-Hall conductivity
\begin{equation}\label{Sinova}
\sigma_{sH}= \frac{e}{8\pi\hbar}
\end{equation}
independent of the Rashba interaction constant $\alpha$ and density $n$,
provided that both spin-split bands are occupied. This is the case when the
density
$n > n^*=m^2\alpha^2/\pi$.

It is important to note that spin current is invariant with respect to time
inversion and thus may exist under equilibrium conditions,
even without any lateral electric field~\cite{Rashba03}. 
For the same reason
spin-Hall conductivity belongs to the family of Fermi-liquid response 
functions,
defined  generally with respect to space-time-inhomogeneous external
electric, $\vec{E} (\Omega,\vec{q})$, and magnetic, 
$\vec{H} (\Omega,\vec{q})$, 
fields.
A Fermi liquid with Rashba spin-orbital coupling
is characterized  by two different spin susceptibilities,
$\chi_{zz}(\Omega,\vec{q})$ and $\chi_{\parallel}(\Omega,\vec{q})$, 
as well as by
the lateral dielectric permeability $\epsilon(\Omega, \vec{q})$. 
Recently, E. Rashba demonstrated~\cite{Rashba_diel} a direct relation 
between the
spin-Hall
conductivity and the dielectric response function
$\epsilon(\Omega, 0)$ of a non-interacting 2DEG with spin-orbital 
interaction.
In this Letter we show that the uniform ($q=0$) spin-Hall conductivity is 
closely related to the in-plane magnetic susceptibility $\chi_\parallel$ 
as well, 
providing additional arguments in favor of the equilibrium nature of 
the spin-Hall constant in a clean 2DEG.

We derive, for a clean (no disorder) 2DEG with an {\em arbitrary 
spin-independent
electron-electron interaction}, the universal relation between 
frequency - dependent
spin-Hall conductivity and Pauli spin susceptibility of 2D electrons
with respect to a spatially uniform parallel magnetic field:
\begin{equation}\label{new}
\sigma_{sH}(\Omega)=\frac{e}{(g\mu_B)^2m_b}\chi_\parallel (\Omega),
\end{equation}
where $m_b$ is the band mass, $\mu_B$ is the Bohr magneton and 
$g$ is the Lande factor. 

The relation (\ref{new}) is valid at any frequency and for
any electron density $n$ consistent with the use of a parabolic band
spectrum, $\epsilon(p) = p^2/2m_b$. This relation (\ref{new})
holds even in the case of very low $n<n^*$  when only one chiral subband is 
populated and the result of Sinova et al.~\cite{Sinova},
Eq.~(\ref{Sinova}), is not applicable.

Next, we calculate corrections to the spin-Hall conductivity from two-particle 
electron-electron
interactions, and find these corrections to be nonzero. 
A direct microscopic calculation 
to the first order in interaction shows
that the electron-electron interaction renormalizes both  
spin-Hall conductivity and in-plane spin susceptibility, 
while keeping relation
(\ref{new}) intact. The relative magnitudes of these corrections are 
proportional to the dimensionless
Coulomb strength $\frac{e^2}{\epsilon \hbar v_F}$ and do not contain 
the spin-orbital subband splitting 
$\Delta$.

Below we provide a brief derivation of the stated results.
We start from the formulation of the model of a two-dimensional electron gas
with Rashba coupling, which is due to the breakdown of 
inversion ("up-down") symmetry, leading to an electric field perpendicular 
to the electron gas plane. It has no effect on the orbital electron 
motion but it does couple to the electron spin via a relativistic 
spin-orbit interaction known as the Rashba term~\cite{Rashba_seminal}. 
The Hamiltonian of an electron consists of the kinetic energy term and 
the Rashba term:
\begin{equation}\label{oneHam}
\hat{h}_{\alpha\beta}(\vec{p})=\frac{p^2}{2m_b}\delta_{\alpha\beta}+
\alpha\left(
\sigma^x_{\alpha\beta} \hat{p}_{y}-\sigma^y_{\alpha\beta}\hat{p}_{x}
\right)  ,
\end{equation}
where $\hat{p}_{\mu}=-i\hbar\partial_{\mu}$ is the momentum of 
the electron, $\alpha$ is the Rashba velocity, $\sigma^i$ ($i=x,y,z$) 
are the Pauli matrices and $\alpha,\beta$ are the spin indices. 
Essential to the following discussion is the parabolic band 
spectrum: $E(p) \propto p^2$.
The Hamiltonian~(\ref{oneHam}) can be diagonalized by the unitary matrix:
\begin{equation}\label{Unitary}
U(\vec{p})={1\over\sqrt{2}} \left( \begin{array}{cc} \displaystyle 1 & 1 
\\ \displaystyle ie^{i\varphi_{\bf p}} & -ie^{i\varphi_{\bf p}} 
\end{array} \right),
\end{equation}
where $\varphi_{\bf p}$ is the angle between the momentum 
$\vec{p}$ of the electron and the $x$-axis, giving the eigenvalues
\begin{equation}\label{eigenv}
E_\lambda(p)=\frac{p^2}{2m_b}-\lambda\alpha p.
\end{equation}
The eigenvalues of the chirality operator, $\lambda=\pm 1$, and
the momentum of the electron
$\vec{p}$ constitute the quantum numbers of an electron state 
$(\vec{p},\lambda)$. The Rashba gas has two 
Fermi circles with the 
different radii: 
$p_F= \sqrt{2m_b\mu+ m_b^2\alpha^2} \pm m_b\alpha$, 
where $\mu$ is the chemical potential. 
We assume that spin-orbital coupling is weak, $\alpha\ll v_F = p_F/m_b$. 
The spin-orbital splitting is then $\Delta=2\alpha p_F$. 
The density of states on the two Fermi circles differs as 
$\nu_{\pm}= \nu (1\pm \alpha/v_F)$, where $\nu=m_b/2\pi\hbar^2$. 
In the following we use units with $\hbar=1$.

We consider the 2D interacting Rashba electron gas at zero 
temperature with the Hamiltonian:
\begin{eqnarray}\label{mainHam}
&&\hat{H}= \int \psi^+_{\alpha}(\vec{r})\hat{h}_{\alpha\beta}(\vec{p})
\psi_{\beta}(\vec{r})\ d^2\vec{r}
\\ \nonumber   {}\\ \nonumber         
&&{}+\frac{1}{2}\int\int
\psi^\dagger_{\alpha}(\vec{r})\psi^\dagger_{\beta}(\vec{r}')
U(|\vec{r}-\vec{r}'|)
\psi_{\beta}(\vec{r}')\psi_{\alpha}(\vec{r})\ d^2\vec{r}d^2\vec{r}',
\end{eqnarray}
where $U(|\vec{r}|)$ is an arbitrary two-electron interaction 
potential; 
$\hat{h}_{\alpha\beta}(\vec{p})$ is defined in 
Eq.~(\ref{oneHam}) and  $\psi^{\dagger}_{\alpha}(\vec{p})$, 
$\psi_{\beta}(\vec{p})$ are the electron creation and 
annihilation operators respectively.
Hamiltonian (\ref{mainHam}) is a rather accurate approximation 
for the clean
two-dimensional 
semiconducting heterostructures.

The electromagnetic vector potential 
$\vec{A}$ couples to the orbital motion of the electron according 
to the transformation: $\vec{p}\rightarrow \vec{p}- e\vec{A}/c$, 
in the Hamiltonian~(\ref{mainHam}). Variation of the 
Hamiltonian~(\ref{mainHam}) 
with respect to $\vec{A}$ gives the electric current operator 
$\hat{J}_{\nu}=e\int \psi^+_{\alpha}(\vec{r})
(\hat{j}_{\nu})_{\alpha\beta} \psi_{\beta}(\vec{r})d^2\vec{r}$, 
where the one-particle current operator is:
\begin{equation}\label{Jx}
(\hat{j}_{\nu})_{\alpha\beta}= 
e\left(\frac{p_{\nu}- \frac{e}{c} A_{\nu}}{m_b}\delta_{\alpha\beta} 
-\epsilon^{\nu i z}\alpha\sigma^i_{\alpha\beta}\right),
\end{equation}
with $\nu=x,y$ being the spatial index and $\epsilon^{zi\mu}$ the 
3D totally antisymmetric tensor. It is actually a velocity 
$\hat{j}_\nu=e\hat{v}_\nu$.

Under the non-uniform SU(2) electron spinor transformation, 
$\psi_\alpha(\vec{r})\mapsto U_{\alpha\beta}(\vec{r}) 
\psi_\beta(\vec{r})$, 
the Hamiltonian~(\ref{mainHam}) becomes dependent on the SU(2) 
``spin electromagnetic'' vector potential 
$\hat{A}_\mu=A^0_\mu\sigma^0+A^i_\mu\sigma^i$, where 
$A^0_\mu$ coincides with the physical electromagnetic potential and 
$A^i_\mu= -i\,\textrm{Tr} (\sigma^iU^+\partial_\mu  U)/2$. Although 
this latter potential is a pure gauge and has no physical consequences, 
variation of the Hamiltonian~(\ref{mainHam}) 
with respect to it defines the spin current of the $i$-th component 
of spin $\frac12$ along the direction $\mu$. The spin current 
$\hat{J}_{\mu}^i= \frac12\int \psi^+_{\alpha}(\vec{r})
(\hat{j}_{\mu}^i)_{\alpha\beta} 
\psi_{\beta}(\vec{r}) d^2\vec{r}$, 
where the single-particle spin current operator reads as
\begin{equation}\label{Jyz}
(\hat{j}_{\mu}^i)_{\alpha\beta}=
\frac12 \left[\frac{p_{\mu}- 
\frac{e}{c}A_{\mu}}{m_b} \sigma^i_{\alpha\beta}+ 
\alpha\epsilon^{i\mu z} \delta_{\alpha\beta} \right].
\end{equation}
Our definition of the spin current~(\ref{Jyz})
is equivalent to
$\hat{J}_{\mu}^i=(\hat{v}_{\mu}
\sigma^i+\sigma^i\hat{v}_{\mu})/4$, cf. 
Refs.~\onlinecite{Murakami,Sinova,MacDonald2,MacDonald3,MacDonald4,
Schliemann,Numeric,loss3,shytov,macdonald5}.

To derive the  relation (\ref{new}) between the spin-Hall constant
and Pauli susceptibility,
we start from two exact commutation relations for total current 
and spin operators.
For the  assumed parabolic band spectrum~(\ref{oneHam}), 
a certain linear combination of the 
total charge current $\vec{J}$ and the total 
spin $\vec{S}$ is
proportional to  the total momentum of the system and commutes 
with the interaction part of the Hamiltonian.
This fact provides us with two exact
commutation relations in the presence of an arbitrary spin-conserving 
two-particle
interaction $U(|\vec{r}-\vec{r}'|)$ in the Hamiltonian~(\ref{mainHam}):
\begin{eqnarray}\label{comm}
&&[\hat{H},\hat{J}_{\mu}]=
-4 i e m_b\alpha^2 \epsilon^{\mu\nu}\hat{J}_{\nu}^z \nonumber \\
\text{and}\nonumber \\
&&[\hat{H},\hat{S}^{\nu}]= 2 i m_b\alpha\hat{J}_{\nu}^z,
\end{eqnarray}
where
$
\hat{S}^{i}=\frac12\int \psi^+_{\alpha}(\vec{r})
\hat{\sigma}^{i}
\psi_{\beta}(\vec{r}) d^2\vec{r}
$
is the total spin of the electron system.

The average spin current of the electron system as a response 
to weak ac electric field, $E_x(t)=E_{0x}\cos\Omega t$, 
is given by the general quantum mechanical
expression
in the first order of perturbation theory~\cite{V}:
\begin{eqnarray}\label{Kubogeneral}
&&\langle \hat{J}_y^z(t)\rangle=\frac{i}{2}
\sum_m\Big[
\left(\hat{J}_x\right)_{m0}
\Big\{\frac{e^{-i\Omega t}}{\Omega (\omega_{m0}-\Omega-i 0)}\\ 
\nonumber
&&{}-\frac{e^{i\Omega t}}{\Omega (\omega_{m0}+\Omega-i0)}\Big\}
\left(\hat{J}_y^z\right)_{0m}-h.c.
\Big]E_{0x},
\end{eqnarray}
where $\omega_{m0}=\epsilon_m-\epsilon_0$, with $0$ being 
the ground state,
and $m$ being the {\em exact} excitation levels of the 
interacting system.
Note that we have used the Kubo formula~(\ref{Kubogeneral})
for external fields homogeneous in space.
 
Using the exact commutation relations (\ref{comm}), 
we can express the matrix elements of the total
charge and spin current operators in the right 
hand side of
Eq.(\ref{Kubogeneral}) in terms of the matrix elements of the total 
spin operator:
\begin{eqnarray}\label{KuboSpin}
\langle \hat{J}_y^z(t)\rangle=-\frac{e}{2m}\sum_m
\Big[
\left(\hat{S}^y\right)_{m0}
\Big\{\frac{e^{-i\Omega t}}{\omega_{mn}-\Omega-i 0}\\ \nonumber
+\frac{e^{i\Omega t}}{\omega_{mn}+\Omega-i 0}\Big\}
\left(\hat{S}^y\right)_{0m}+
h.c.\Big]E_{0x}.
\end{eqnarray}

Note that now the right hand side of Eq.~(\ref{KuboSpin})
is fully analogous (up to a replacement of the $e/m_b$ factor by 
$(g\mu_b)^2$) to the linear-response expression for Pauli
spin susceptibility with respect to an in-plane magnetic field
$H_y(t)=H_{0y}\cos\Omega t$,
which would replace the electric field $E_{0x}$.
This observation leads us immediately to the relation (\ref{new})
which is the main result of the present Letter.
This relation holds, remarkably, for linear response to 
perturbations
with an arbitrary frequency $\Omega$ which are uniform 
in space, i.e. $q=0$.

The Fermi liquid response function usually depends on the ratio 
$\omega/qv_F$.
For example, in a normal isotropic Fermi liquid $\chi=0$ if the 
limit $q \to 0$, 
$\omega \to 0$
is taken with the ratio $q v_F/\omega \to 0$ as a consequence of 
the total
spin conservation. The standard Pauli susceptibility 
$\chi_{Pauli}=2\mu_B^2\nu(\epsilon_F)$ is obtained with the 
opposite order of 
limits,
$\omega/qv_F \to 0$ and $q\rightarrow 0$.
In the case of the Rashba Fermi gas Gor'kov and
Rashba~\cite{GR} have found that
$\chi_{zz}=\chi_{\parallel}=\chi_{Pauli}$ at
$\omega/qv_F=0$ and $q\rightarrow 0$. We find
$\chi_{\parallel}=\frac{1}{2}\chi_{Pauli}$ and
$\chi_{zz}=\chi_{Pauli}$ at $q v_F/\omega=0$ and
$\omega\rightarrow 0$. Therefore we expect the relation (\ref{new})
to be valid for $ qv_F /\omega \ll 1$.

\begin{figure}
\includegraphics[angle=0,width=0.25\textwidth]{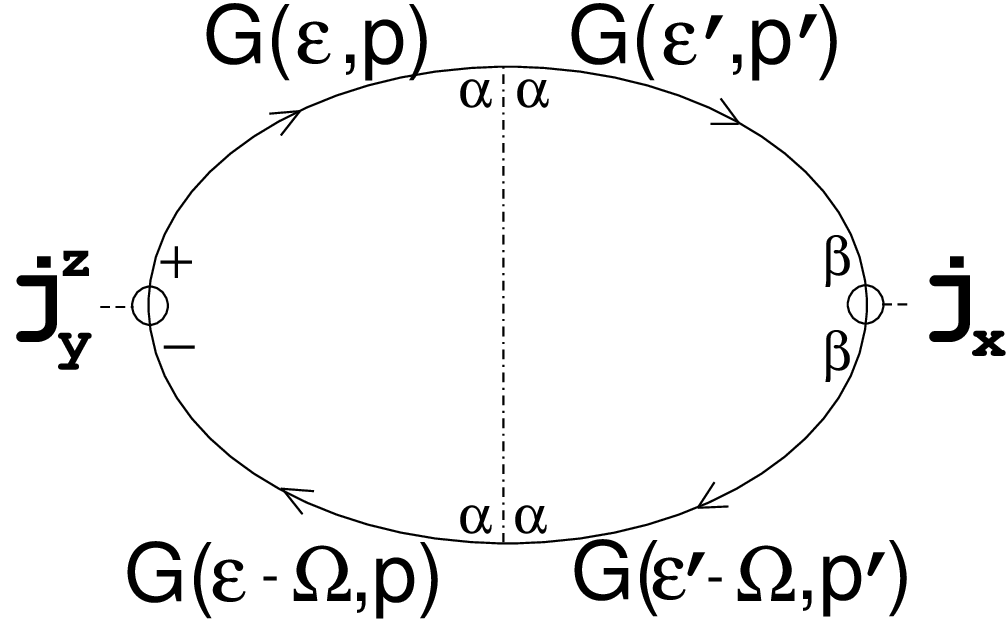}
\vskip 0.545in
\includegraphics[angle=0,width=0.25\textwidth]{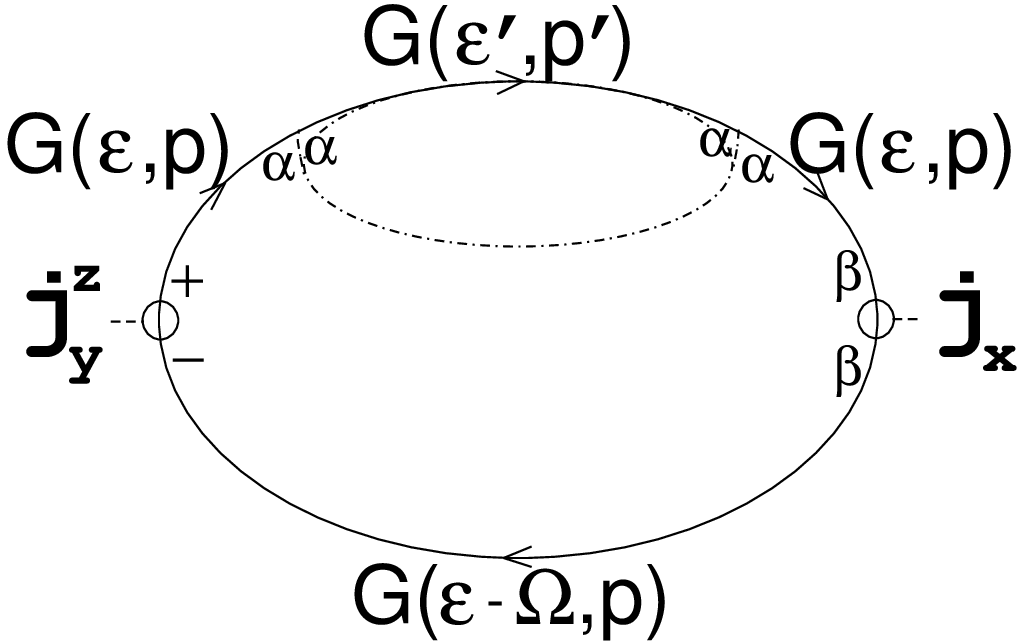}
\vskip 0.545in
\includegraphics[angle=0,width=0.25\textwidth]{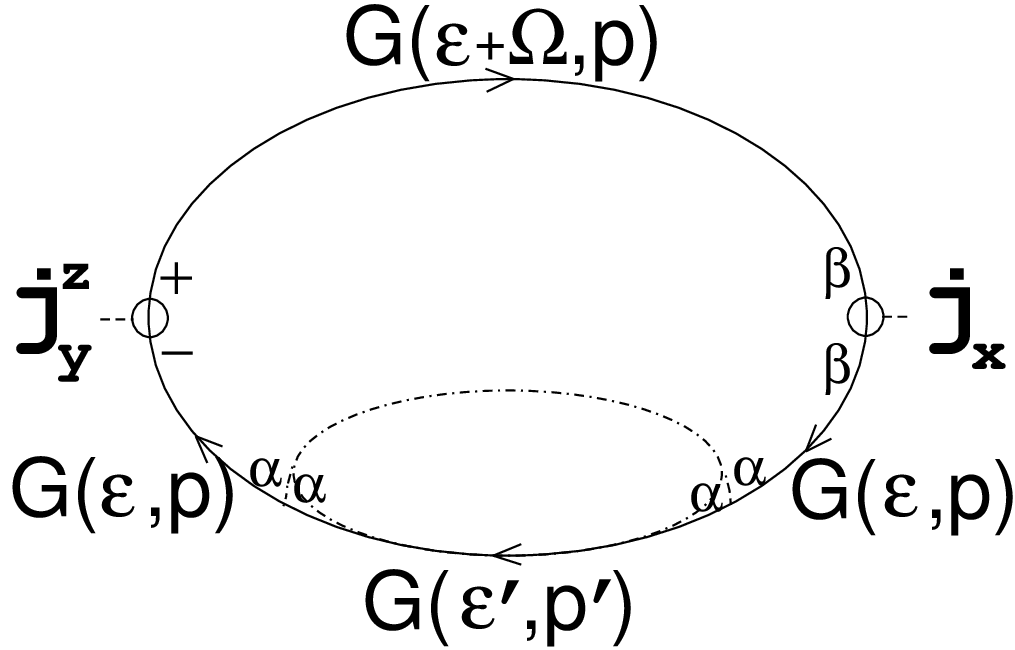}
\caption{\label{loop} The correction to the spin-Hall conductivity 
from electron-electron interaction is given by the sum of the 
three diagrams, which have equal sign and coefficient.
Indices $+,-,\alpha,\beta$ correspond to Keldysh space.
Dashed lines correspond to interaction
$U(|\vec{p}-\vec{p}'|)$.
}
\end{figure}

Next we switch to the second subject of this Letter, which is
a calculation of interaction corrections to the spin-Hall 
conductivity~(\ref{Sinova}).
We use the Keldysh technique~\cite{Keldysh}.
In the lowest order of the  e-e interaction $U(\vec{r})$, three diagrams,
shown in Fig.\ref{loop}, contribute to $\delta\sigma_{sH}$.

Our result is given by Eq.~(\ref{ABC}).

The averaged Keldysh Green's function is a four by four matrix 
$\mathcal{G}(p,\epsilon)$ that can be conveniently factorized 
into a two by two Keldysh matrix whose elements are matrices 
in spin space:
\begin{eqnarray}\label{G_Keldysh}
&&\left(
\begin{array}{cc} \mathcal{G}_{--} & \mathcal{G}_{-+} \\ 
\mathcal{G}_{+-} & \mathcal{G}_{++} \end{array} 
\right)= \left(
\begin{array}{cc} 1-N(p) & -N(p)\\ 
1-N(p) & -N(p) \end{array} 
\right) G^R(p,\epsilon)+ \nonumber\\ 
\nonumber \\
&&+\left(\begin{array}{cc} N(p) & N(p)\\ 
-1+N(p)&-1+N(p) \end{array} 
\right)G^A(p,\epsilon), 
\end{eqnarray}
where the electron distribution function $N(p)$ is 
a matrix in spin space.
The retarded and advanced averaged Green's functions 
are diagonal in the chiral basis: 
$G^{(R,A)}_{\lambda'\lambda}(\epsilon,\vec{p})= 
G^{(R,A)}_{\lambda}(\epsilon,\vec{p}) \delta_{\lambda'\lambda}$, 
and the solution to the Dyson equation reads~\cite{AGD}:
\begin{equation}\label{Greenfunction}
G^{R,A}_{\lambda}(\epsilon,\vec{p})= 
\frac{1}{\epsilon-\epsilon_\lambda(\vec{p}) +\mu \pm i0}
\delta_{\lambda'\lambda}.
\end{equation}

We choose the gauge for the uniform electric field 
$\vec{E}(t)=\vec{E}(\Omega) e^{-i\Omega t}$ to be a 
time dependent vector potential 
$\vec{A}(t)=\vec{A}(\Omega) e^{-i\Omega t}$, 
where $\vec{A}(\Omega)= -ic\vec{E}(\Omega)/\Omega$. 
Using the Keldysh technique we average 
the spin current operator over the electron state perturbed 
by both the electromagnetic Hamiltonian, 
$\hat{H}_{em}=-\frac{1}{c} \int d^2\vec{r} 
\hat{j}_{\nu}(\vec{r}) A_{\nu}(t)$, and the electron-electron interaction
Hamiltonian, $\frac{1}{2}\int\int
\psi^\dagger_{\alpha}(\vec{r})\psi^\dagger_{\beta}(\vec{r}')
U(|\vec{r}-\vec{r}'|)
\psi_{\beta}(\vec{r}')\psi_{\alpha}(\vec{r})\ d^2\vec{r}d^2\vec{r}'$, 
to the first order of perturbation theory. The correction to the 
spin-Hall 
conductivity $\delta\sigma_{sH}$ is then found from the 
relationship $\langle\hat{j}_{\mu}^z(\Omega)\rangle= 
\epsilon_{\mu\nu} (\sigma_{sH}(\Omega) +
\delta\sigma_{sH}(\Omega)) E_{\nu}(\Omega)$.
The resulting expression for the correction to the spin-Hall 
conductivity from electron-electron interactions reads as follows:
\begin{eqnarray}\label{ABC}
&&\delta\sigma_{sH}=\frac{e}{2V \Omega}\sum_{p,p'}\int\frac{d\epsilon}{2\pi}
\frac{d\epsilon'}{2\pi}
\text{Tr}\left[A+B+C\right]_{-+}U(|\vec{p}-\vec{p}'|),\nonumber \\
&&\text{where}\nonumber \\
&&A=J^z_y\mathcal{G}(\epsilon-\Omega,p)\left[\tau_z,
\mathcal{G}(\epsilon'-\Omega,p')\tau_zJ_x\mathcal{G}(\epsilon',p')\right]_+
\mathcal{G}(\epsilon,p),\nonumber \\
&&B=J^z_y\mathcal{G}(\epsilon-\Omega,p)\tau_zJ_x
\mathcal{G}(\epsilon,p)\left[\tau_z,\mathcal{G}(\epsilon',p')\right]_+
\mathcal{G}(\epsilon,p),\nonumber \\
&&C=J^z_y\mathcal{G}(\epsilon,p)
\left[\tau_z,\mathcal{G}(\epsilon',p')\right]_+
\mathcal{G}(\epsilon,p)\tau_zJ_x\mathcal{G}(\epsilon+\Omega,p),\nonumber \\
\end{eqnarray} 
$\tau^z$ is the four by four matrix 
given by the direct 
product of the Pauli matrix $\sigma^z$ in the Keldysh 
space and the unit matrix in the spin space. The current 
operators in Eq.~(\ref{ABC}) are the direct 
products of matrices~(\ref{Jx},~\ref{Jyz}) and the unitary 
matrix in the Keldysh space. The $\textrm{Tr}$ in 
Eq.~(\ref{ABC}) operates only in the spin 
space whereas the indices $-+$ correspond to 
the Keldysh space. 

Taking the $-+$ element in Keldysh space in Eq.~(\ref{ABC}), 
using (\ref{G_Keldysh}),
then taking the trace in spin space and performing integration 
over the energies,
we obtain the expression for the correction to the spin-Hall 
conductivity 
in the lowest order of the electron-electron interaction:
\begin{eqnarray}\label{answer}
&&\delta\sigma_{sH}=\nonumber \\
&&-e\int\int \frac{d^2 \vec{p}}{(2\pi)^2} \frac{d^2 \vec{p}'}{(2\pi)^2}
\delta N(p)\delta N(p')\, 
U(|\vec{p}-\vec{p}'|) F(\vec{p},\vec{p}').\nonumber \\ 
&&\text{Here}\nonumber \\ 
&&\delta N(p)=N_+(p)-N_-(p)\text{ and}\nonumber \\ 
&&F(\vec{p},\vec{p}')=\frac{\cos{(\varphi-\varphi')}
\left\{-p^2+pp'\cos{(\varphi-\varphi')}
\right\}}{16 m_b \alpha^2 p^2p'^2},\nonumber \\ 
\end{eqnarray}
with $N_{\pm}(p)$ being the distribution functions 
of the two Fermi circles of different chiralities. For zero 
temperature $N_{\pm}(p)=\theta(-p+p_{F\pm}).$

Explicit integration over momenta in the expression (\ref{answer}) 
was performed  for small
spin-orbit interaction $\alpha/v_F \ll 1$ in two limiting cases:  
short-ranged two-particle interaction (Coulomb potential screened on the
lengthscale $\kappa^{-1}$ smaller than interparticle distance), and full
 long-range Coulomb interaction.
In the Fourrier space these interaction potentials are: 
$U_1(|\vec{p}-\vec{p}'|)=\frac{2\pi e^2}{\kappa\epsilon}$, and
$U_2(|\vec{p}-\vec{p}'|) = \frac{2\pi e^2}{\epsilon |\vec{p}-\vec{p}'|}$.
The final expressions for $\sigma_{sH}$ in these two cases are as follows:
\begin{equation}
\sigma_{sH}^{(\rm short)}=\frac{e}{8\pi\hbar}\left[1-\frac{m_be^2}
{2 \epsilon \kappa}\right]
\label{corr1}
\end{equation}
for the short-range potential and
\begin{equation}
\sigma_{sH}^{(\rm Coulomb)}=\frac{e}{8\pi\hbar}\left[1-\frac{2 m_b
e^2}{3 \pi \epsilon p_F}\right]
\label{corr2}
\end{equation}
for the Coulomb potential. 
It is seen that the correction to the spin-Hall conductivity is 
independent of
the spin-orbit constant $\alpha$ (in Eq.~(\ref{corr2}) corrections of the
order $(\alpha/v_F)^2 \ll 1$ are neglected),
and is proportional to the standard Coulomb interaction parameter 
$e^2/\epsilon\hbar v_F$.

For completeness we have performed direct diagram calculations of
the interaction correction to the in-plane susceptibility,
represented by three diagrams similar to those shown in Fig.1.
The results for the relative corrections to the in-plane susceptibility
were found to coincide with expressions (\ref{corr1},\ref{corr2}),
in  agreement with the general relation (\ref{new}).

We checked by direct calculation for a clean system without interaction
that spin susceptibilities and spin-Hall conductivities, Eq.~(\ref{sHj/2}), 
for systems of fermions
with higher spins $j$ follow relation (\ref{new}).
Interestingly we find \cite{od} that in the case of an ideal 2D Rashba gas 
of fermions 
of arbitrary half-integer spin $j$ the value of the spin-Hall constant 
is also universal
and grows
with $j$:
\begin{eqnarray}\label{sHj/2}
\sigma_{sH}(j)=\frac{e}{4\pi}\sum_{m=-j}^{j}m^2.
\end{eqnarray} 

In this Letter we did not discuss the very actively 
debated issue of 
the stability of the
spin-Hall response to disorder, with quite a few of 
conflicting results
presented during last months~\cite{MacDonald4,
Schliemann,Molenkamp2,Numeric,shytov,macdonald5,loss3}.
We expect our results to be directly relevant
for submicron samples of a very clean electron gas, with the 
sample size 
less than the
elastic scattering length. The influence of disorder upon
$\sigma_{sH}$ in the presense of electron-electron interactions 
is to be studied separately. 

In conclusion, we have shown that the frequency-dependent spin-Hall 
conductivity and Pauli
susceptibility of a clean interacting 2D Rashba EG are 
proportional to each
other, with the coefficient containing band mass, Lande factor 
and Bohr magneton only.
We calculated the first-order interaction-induced correction to 
the spin-Hall conductivity
and found it to be proportional to the standard dimensionless 
interaction strength.
At the final stage of preparation of this Letter, we became aware 
of the paper~\cite{Loss}, where a similar relation between spin-Hall 
conductivity 
and Pauli susceptibility 
is discussed for a non-interacting Rashba electron gas.

The author thanks M. Feigel'man for discussions and N. Sedlmayr 
for reading the paper. This research was supported by the Dynasty Foundation,  
Landau Scholarship-Juelich,
Program "Quantum Macrophysics" of Russian Academy of Sciences, and the
RFBR grant \# 04-02-16348.

\end{document}